\documentclass[a4paper,11pt]{article}
\pdfoutput=1 

\usepackage{jheppub} 
\usepackage[T1]{fontenc} 

\usepackage{slashed}

\newcommand{\ri}{\mathrm{i}}

\title{\boldmath Holographic Wilson Loop One-point Functions in ABJM Theory}

\author[a]{Xiao-Yi Zhang}
\author[b,c]{Yunfeng Jiang}
\author[a,c]{Jun-Bao Wu\footnote{ \textit{Note:
The unusual ordering of authors instead of the standard alphabetical one in hep-th community is for
students to get proper recognition of contribution under the current out-dated practice in China.}}}

\affiliation[a]{Center for Joint Quantum Studies and Department of Physics, School of Science, Tianjin University, 135 Yaguan Road, Tianjin 300350, P. R. China}
\affiliation[b]{School of physics and Shing-Tung Yau Center, Southeast University, Nanjing 211189, P. R. China}
\affiliation[c]{Peng Huanwu Center for Fundamental Theory, 96 Jinzhai Road, Hefei, Anhui 230026. P. R. China}
\preprint{USTC-ICTS/PCFT-25-31}
\emailAdd{xiaoyi\_zhang@tju.edu.cn}
\emailAdd{jinagyf2008@seu.edu.cn}
\emailAdd{junbao.wu@tju.edu.cn}

\abstract{We compute the correlation function between a circular half-BPS Wilson loop (or straight Wilson line) and a local operator in ABJM theory utilizing its M-theory description. The local operator can be a $1/3$-BPS single-trace chiral primary operator or the stress-energy tensor. Using the AdS/CFT correspondence, these correlators are dual to fluctuations of a probe M2-brane in $AdS_4 \times S^7/\mathbb{Z}_k$. We derive analytic results for both cases and compare them with existing results based on supersymmetric localization in the literature. In the large-$N$ limit with $k$ finite, our holograkphic results exhibit perfect agreement with localization. }

\begin{document} 
\maketitle

\flushbottom

\section{Introduction}
\label{sec:intro}
Conformal field theories (CFTs) can be enriched by the inclusion of extended objects such as line and surface defects. Among these, Wilson lines and loops (WLs) have been the most extensively studied due to their rich physical and mathematical properties. In supersymmetric quantum field theories (SQFTs), certain classes of WLs preserve enough supersymmetry to allow for exact computations, making them invaluable probes of non-perturbative dynamics. The study of WLs has led to significant advances through the interplay of diverse non-perturbative methods, including holography~\cite{Maldacena:1998im, Rey:1998ik, Drukker:2005kx, Yamaguchi:2006tq, Dymarsky:2006ve, DrukKer:2006ga, Drukker:2006zk, Drukker:2007dw, Drukker:2007yx, Drukker:2007qr}, supersymmetric localization~\cite{Pestun:2007rz, Kapustin:2009kz, Pestun:2009nn, Giombi:2009ds, Giombi:2012ep, Giombi:2018qox}
the conformal bootstrap~\cite{Liendo:2016ymz, Liendo:2018ukf, Barrat:2021yvp, Barrat:2022psm,  Bianchi:2020hsz}, integrability~\cite{Drukker:2005cu, Drukker:2006xg, Correa:2012hh, Gromov:2015dfa, Gromov:2016rrp, Cavaglia:2018lxi, Gromov:2021ahm, Correa:2018fgz, Giombi:2018qox, Grabner:2020nis,  Cavaglia:2022yvv, Correa:2023lsm, Jiang:2023cdm},  and \textit{bootstrability} (the combination of the last two tools)~\cite{Cavaglia:2021bnz, Cavaglia:2022qpg, Cavaglia:2024dkk}. Moreover, WLs play a central role in the AdS/CFT correspondence, where they correspond to fundamental strings~\cite{Maldacena:1998im, Rey:1998ik}/D-branes~\cite{Drukker:2005kx, Yamaguchi:2006tq} in the bulk or certain bubbling geometries~\cite{Yamaguchi:2006te, Lunin:2006xr, DHoker:2007mci}, thereby enriching the holographic dictionary~\cite{Maldacena:1997re, Gubser:1998bc, Witten:1998qj}.\par

The investigation of WLs in $\mathcal{N}=4$ super-Yang-Mills (SYM) theory has a long and fruitful history. A non-perturbative result in terms of matrix model integral was proposed based on the structure of perturbative results~\cite{Erickson:2000af}, confirming previous holographic calculations at strong coupling~\cite{ Berenstein:1998ij, Drukker:1999zq}. A major breakthrough came with the application of supersymmetric localization, which reduced the computation of WL expectation values to a matrix model~\cite{Pestun:2007rz}, which is the same one conjectured from the perturbative computations. This framework was subsequently extended to less supersymmetric WLs~\cite{Zarembo:2002an, Dymarsky:2006ve, DrukKer:2006ga, Drukker:2006zk, Drukker:2007dw, Drukker:2007yx, Drukker:2007qr, Pestun:2009nn}. In contrast, WLs in the three-dimensional $\mathcal{N}=6$ ABJM theory~\cite{Aharony:2008ug} have received comparatively less attention. So far, the focus has been on the bosonic 1/6-BPS loop~\cite{Drukker:2008zx, Chen:2008bp, Rey:2008bh} and the 1/2-BPS loops~\cite{Drukker:2009hy}~\footnote{A nice review on various aspects of BPS WLs in ABJM theory is~\cite{Drukker:2019bev}.}. The localization of these WLs leads to a highly non-trivial matrix models~\cite{Kapustin:2009kz}, which are much more challenging to evaluate and thus spurred the development of the Fermi gas approach to evaluate the resulting matrix integrals~\cite{Marino:2011eh, Klemm:2012ii}. The holographic duals of some WLs have been explored in~\cite{Drukker:2008zx, Rey:2008bh}, where they were identified with (smeared) string and D-brane configurations in $AdS_4 \times \mathbb{CP}^3$, or (smeared) membrane and other objects in $AdS_4\times S^7/{\mathbb{Z}_k}$. We recall that ABJM theory has two kinds of large N limit,
 \begin{itemize}
    \item 't~Hooft limit: take $N, k\to \infty$ with the 't~Hooft coupling $\lambda=N/k$ fixed. Large $\lambda$ limit in  this case is dual to type IIA superstring theory on $AdS_4 \times \mathbb{CP}^3$.
    \item M-theory limit: take $N\to \infty$ with $k$ kept finite. The holographic description is in terms of M-theory on $AdS_4\times S^7/\mathbb{Z}_k$.
\end{itemize}\par

A CFT coupled to extended objects is referred to as a defect CFT (dCFT), where the defect introduces additional observables, such as correlation functions in the presence of the defect. Among the simplest yet most important observables in a dCFT are defect one-point functions, which encode the operator product expansion (OPE) coefficients of the WL. Physically, these coefficients describe how a small WL can be approximated by a local operator insertion, making them fundamental data for characterizing the defect~\cite{Berenstein:1998ij, Billo:2016cpy}. While such OPE coefficients are generally challenging to compute, progress has been made in certain cases using bootstrap techniques~\cite{Billo:2016cpy, Gadde:2016fbj}, integrability~\cite{Jiang:2023cdm}, localization~\cite{Giombi:2009ds, Giombi:2012ep}, and holography~\cite{Berenstein:1998ij, Semenoff:2006am, Giombi:2006de}.\par

In this paper, we study WL one-point functions in ABJM theory, focusing on the cases where the WL is either an infinite straight line or a circular loop. The local operators under consideration are BPS operators: the stress-energy tensor and a 1/3-BPS chiral primary operator (CPO). We compute these correlation functions holographically by evaluating specific fluctuations of the classical membrane configuration in $AdS_4 \times S^7/\mathbb{Z}_k$. Beyond their intrinsic physical interest, such OPE coefficients in the M-theory limit remain scarce in the literature, making our results a valuable addition to the strong coupling data for studying defect CFTs. Moreover, our approach provides a solid basis for computing similar observables of other types of defect correlation functions in ABJM theory.\par

To validate our holographic results, we compare them with existing localization computations for these correlators~\cite{Lewkowycz:2013laa, Guerrini:2023rdw, Beccaria:2025vdj}. By taking the M-theory limit, we find agreement between the two approaches. In the case of CPO, the localization result is only available for the operator with conformal dimension $\Delta=1$, which we compared with the holographic result and find a perfect match. Notably, the matching requires careful normalization of the OPE coefficients, particularly for the CPO, where the two-point function normalization is a highly non-trivial function of the gauge group rank $N$ and the Chern-Simons level $k$ (Two-loops correction to this normalization factor in the field theory side were computed in~\cite{Young:2014lka, Young:2014sia}).\par

The remainder of this paper is organized as follows. In section~\ref{sec:corr1}, we outline the general setup for WL one-point functions, detailing the relevant operators and the constraints imposed by conformal symmetry. Conformal invariance fixes the spacetime dependence of these correlators, reducing the problem to the computation of a single OPE coefficient. In section~\ref{sec:CorrHolo}, we present the holographic computation of these coefficients. In section~\ref{sec:localization}, we compare our results with localization predictions, verifying their consistency in the M-theory limit. We conclude in section~\ref{sec:concl} with a discussion of future directions.

\section{Correlators of a BPS Wilson loop and a local operator }
\label{sec:corr1}
In this section, we outline the setup for computing the correlation functions of interest in ABJM theory, which is a three-dimensional $\mathcal{N}=6$ super-Chern-Simons theory with gauge group $U(N)_k \times U(N)_{-k}$. The theory contains two gauge fields $A_\mu$ and $\hat{A}_\mu$, together with four scalars $Y^I$ and four Dirac spinors $\psi_I$ in the bi-fundamental representation of the gauge group. We work in Euclidean $\mathbb{R}^3$ with coordinates $x^\mu = (x^1, x^2, x^3)$~\footnote{We use $\mu, \nu, \cdots$ to denote  coordinate of $ \mathbb{R}^3$, $M, N, \cdots$ coordinates of $AdS_4\times S^7/ \mathbb{Z}_k$, $m, n, \cdots$ coordinates of $AdS_4$,  $\alpha, \beta, \cdots$ coordinates of $S^7/ \mathbb{Z}_k$, and finally $a, b, \cdots$ coordinates of the worldvolume of probe M2-brane.}.


\paragraph{Half-BPS Wilson line/loop} We consider a half-BPS Wilson line along the $x^1$-axis, localized at $x^2 = x^3 = 0$. The contour $L$ is parameterized as
\begin{equation}
x^\mu(\tau)=(\tau, 0, 0)\,,\, -\infty<\tau<\infty\,.
\end{equation}
The Wilson line in the fundamental representation of the supergroup $U(N|N)$ is given by
\begin{equation}\label{WL}
W[L]=\mathrm{Tr} \mathcal{P}\, \exp \left(-{\mathrm{i}} \int_L d\tau \mathcal{L}_{1/2}^L(\tau)  \right)\,,
\end{equation}
where the superconnection $\mathcal{L}_{1/2}^L$ takes the form
\begin{equation}
\mathcal{L}_{1/2}^{L}=\left(
\begin{array}{cc}
\mathcal{A}^L & \bar{f}_1^L\\
f_2^L         &  \hat{\mathcal{A}}^L
\end{array}
\right)\,.
\end{equation}
The components are explicitly:
\begin{eqnarray}
    \mathcal{A}^L&=&A_\mu \dot{x}^\mu+\frac{2\pi}k (\delta^J_I-2 \alpha_I\bar{\alpha}^J)Y^I\bar{Y}_J|\dot{x}|\,,\\
    \hat{\mathcal{A}}^L&=&\hat{A}_\mu \dot{x}^\mu+\frac{2\pi}k (\delta^J_I-2\alpha_I \bar{\alpha}^J)\bar{Y}_JY^I|\dot{x}|\,,\\
    \bar{f}_1^L&=&\mathrm{i}\sqrt{\frac{2\pi}{k}}\bar{\alpha}^Iu_+\psi_I|\dot{x}|\,,\\
    f_2^L&=&\mathrm{i} \sqrt{\frac{2\pi}{k}}\bar{\psi}^Iu_-\alpha_I|\dot{x}|\,.
\end{eqnarray}
Here, $\dot{x}^\mu=dx^\mu/d\tau$, and  $\alpha_I$ are complex parameters satisfying $\sum_{I=1}^4 \alpha_I \bar{\alpha}^I = 1$, with $\bar{\alpha}^I = (\alpha_I)^*$. The spinors $u_\pm$ follow the conventions of \cite{Mauri:2018fsf} and are given by
\begin{equation}\label{eq:spinors}
 u_{\pm}^\alpha=(\mp \mathrm{i}, -1)\,, \quad u_{\pm\alpha}=\left(\begin{array}{c} 1 \\ \mp \mathrm{i}\end{array}\right)\,.
\end{equation}

We also consider the following half-BPS circular Wilson loop along the circle $C=\{x^\mu=(R\cos\tau, R\sin\tau, 0)|\tau\in [0, 2\pi]\}$,
\begin{equation}
W[C]=\mathrm{Tr} \mathcal{P}\, \exp \left(-\mathrm{i} \oint_C d\tau \mathcal{L}_{1/2}^C(\tau)  \right)\,,
\end{equation}
with the superconnection given by
\begin{equation}
\mathcal{L}_{1/2}^{C}=\left(
\begin{array}{cc}
\mathcal{A}^C & \bar{f}_1^C\\
f_2^C         &  \hat{\mathcal{A}}^C
\end{array}
\right)\,.
\end{equation}
The components explicitly read:
\begin{eqnarray}
\mathcal{A}^C&=&A_\mu \dot{x}^\mu-\mathrm{i}\frac{2\pi}{k}(2\alpha_I\bar{\alpha}^J-\delta_I^J)|\dot{x}|Y^I\bar{Y}_J\,,\\ \label{eq:line1}
\hat{\mathcal{A}}^C&=&\hat{A}_\mu \dot{x}^\mu-\mathrm{i}\frac{2\pi}{k}(2\alpha_I\bar{\alpha}^J-\delta_I^J)|\dot{x}|\bar{Y}_JY^I\,,\\ \label{eq:line2}
\bar{f}_1^C&=&-\sqrt{\frac{2\pi}{k}}\bar{\alpha}^I\bar{\zeta}\psi_I|\dot{x}|\,,\\ \label{eq:line3}
f_2^C&=&\sqrt{\frac{2\pi}{k}}\bar{\psi}^I \eta \alpha_I |\dot{x}|\,,\label{eq:line4}
\end{eqnarray}
where $\alpha_I$ and $\bar{\alpha}^I$ are the same as the Wilson line case, and the spinors $\bar{\zeta}^\alpha$ and $\eta_\alpha$
are
\begin{eqnarray}
\bar{\zeta}^\alpha=\left(e^{\mathrm{i}\tau/2}, e^{-\mathrm{i}\tau/2}\right)\,,\quad \eta_\alpha=\left(\begin{array}{c}
e^{-\mathrm{i}\tau/2}\\
e^{\mathrm{i}\tau/2}
\end{array}\right)\,.
\end{eqnarray}

\paragraph{Local operators} We are interested in the correlation functions of the Wilson line/loop and a local operator. In this work, we will consider two types of local operators: 1) The stress energy tensor of the theory; 2) The $1/3$-BPS chiral primary operator.

Our choice of the normalization of the stress energy tensor follows the one in~\cite{Lewkowycz:2013laa}, 
\begin{equation}
    T_{\mu\nu}=-\frac{2}{\sqrt{g}}\frac{\partial (\sqrt{g}\mathcal{L})}{\partial g^{\mu\nu}}\Bigg{|}_{g_{\mu\nu}=\delta_{\mu\nu}}\,.
\end{equation}
where $\mathcal{L}$ is the Lagrangian density of the theory.
The definition of the $1/3$-BPS  CPO is
\begin{equation}\label{operator}
\mathcal{O}^A=(C^A)^{J_1\cdots J_L}_{I_1\cdots I_L} \mathrm{Tr}(Y^{I_1}\bar{Y}_{J_1}\cdots Y^{I_L}\bar{Y}_{J_L})\,,
\end{equation}
with $C^A$ being a symmetric traceless tensor, satisfying
\begin{equation}(C^A)^{J_1\cdots J_L}_{I_1\cdots I_L}=(C^A)^{J_1\cdots J_L}_{(I_1\cdots I_L)}=(C^A)^{(J_1\cdots J_L)}_{I_1\cdots I_L}\,
\end{equation}
and
\begin{equation}
\delta^{I_1}_{J_1}(C^A)^{J_1\cdots J_L}_{I_1\cdots I_L}=0\,.
\end{equation}
We also demand that $C^A$ is normalized
\begin{equation}
(C^A)^{J_1\cdots J_L}_{I_1\cdots I_L}(\bar{C}^B)^{I_1\cdots I_L}_{J_1\cdots J_L}= \delta^{AB}\,,
\end{equation}
where $\bar{C}^B$ is the complex conjugate of $C^B$.\par

\begin{figure}[h!]
\centering
\includegraphics[scale=0.5]{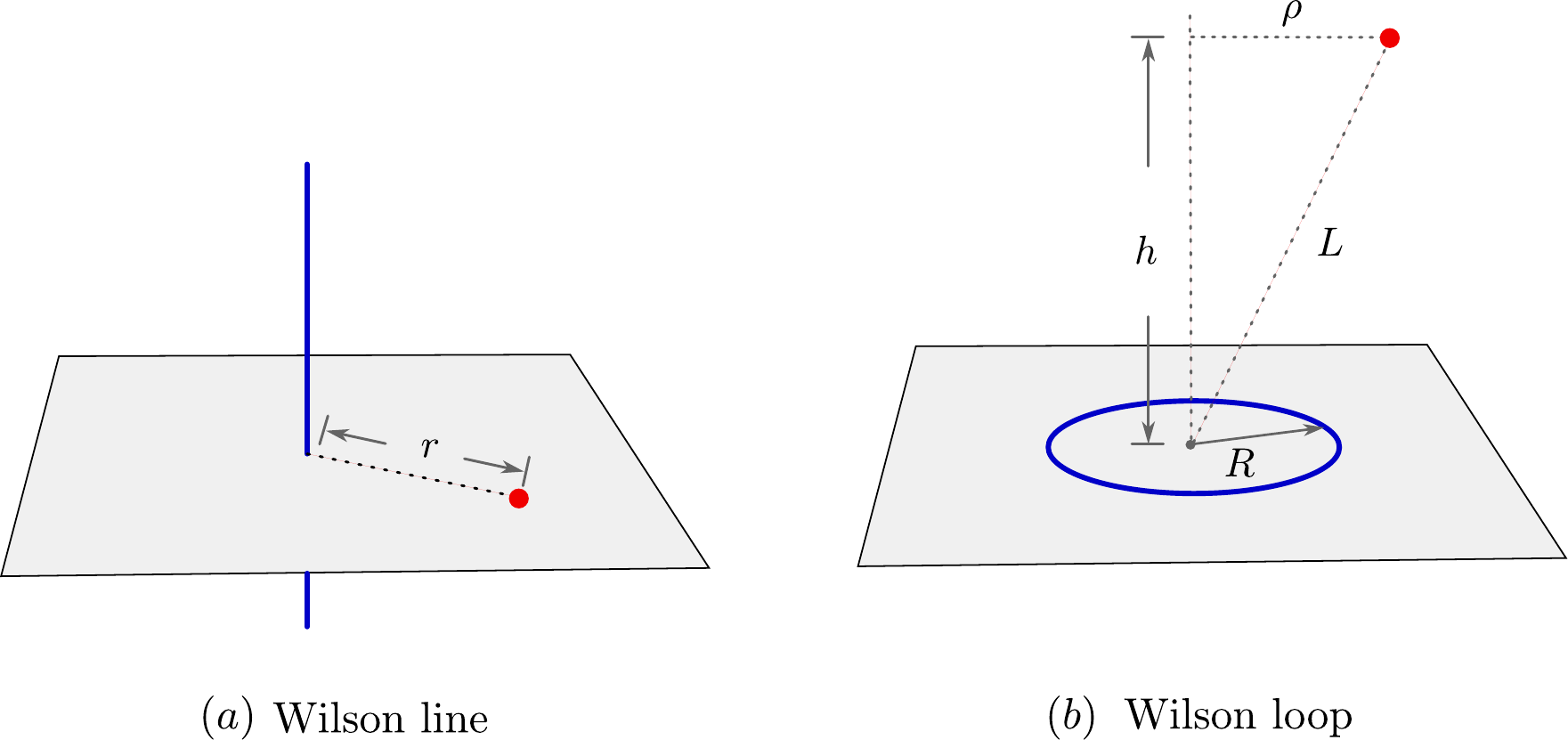}
\caption{Correlation function of a Wilson line/loop and a local operator. The blue line in the left and right panel stands for a Wilson line and Wilson loop, respectively. The red bullet denote a local operator. In the case of the Wilson loop, we consider the OPE limit $L\gg R$.}
\label{fig:WL}
\end{figure}

For the case of a Wilson line, we place the local operator away from the line, as is shown in the left panel of Figure~\ref{fig:WL}. For the case of a Wilson loop, we place the operator far away from the Wilson loop, which is also known as the OPE limit. This is shown in the right panel of Figure~\ref{fig:WL}.

\paragraph{OPE coefficients} The spacetime dependence of the correlation function is fixed by conformal symmetry up to a constant. For the Wilson line one-point function of a CPO, we have 
\begin{align}
\frac{\langle W[L]\mathcal{O}(r)\rangle}{\langle W[L]\rangle}=\frac{B_\mathcal{O}}{r^{\Delta}}
\end{align}
where $r$ is the distance between the local operator and the Wilson line and $\Delta$ is the scaling dimension of the local operator. We will compute the coefficient $B_\mathcal{O}$. In order to compare the results from localization and  holography, we  further divide $B_\mathcal{O}$ by a proper normalization of the two-point function and consider
\begin{align}
\frac{\langle W[L]\mathcal{O}(r)\rangle}{\sqrt{\mathcal{N}_{\mathcal{O}}}\langle W[L]\rangle}
=\frac{B_\mathcal{O}}{\sqrt{\mathcal{N}_{\mathcal{O}}}}\frac{1}{r^{\Delta}}\,.
\end{align}
For the circular Wilson loop with contour $x(\tau)=(R\cos\tau,R\sin\tau,0)$ and a local operator at a generic point $(y_1,y_2,y_3)$, conformal symmetry fixes the correlation function to be of the form~\cite{Berenstein:1998ij, Alday:2011pf}
\begin{align}\label{eq:setupWO}
\frac{\langle W[C]\mathcal{O}(y)\rangle}{\sqrt{\mathcal{N}_{\mathcal{O}}}\langle W[C]\rangle}
=C_{\mathcal{O}}\frac{R^{\Delta}}{\left((h^2+\rho^2-R^2)^2+4R^2h^2\right)^{\Delta/2}}
\end{align}
where  $\rho=\sqrt{y_1^2+y_2^2}$, $h=|y_3|$ and $C_{\mathcal{O}}$ is a constant. In the OPE limit $L=\sqrt{\rho^2+h^2}\gg R$, we have
\begin{align}
\frac{\langle W[C]\mathcal{O}(y)\rangle}{\sqrt{\mathcal{N}_{\mathcal{O}}}\langle W[C]\rangle}=C_{\mathcal{O}}\frac{R^{\Delta}}{L^{2\Delta}}
\end{align}
The coefficients $B_{\mathcal{O}}$ and $C_{\mathcal{O}}$ are related by~\cite{Lewkowycz:2013laa}
\begin{equation}
    \frac{B_{\mathcal{O}}}{\sqrt{\mathcal{N}_\mathcal{O}}}=\frac{C_{\mathcal{O}}}{2^\Delta}\,.
\end{equation}
We will review the derivation of this relation in subsection~\ref{subsection:fieldWO}.

For a Wilson line $W[L]$ at $x_2=x_3=0$ (extended in $x_1$ direction) and stress tensor $T_{\mu\nu}$ located at $(y_1, y_2, y_3)$~\cite{Kapustin:2005py}, conformal symmetry fixes the correlation function of different components of the stress tensor to be
\begin{eqnarray}
\label{eq:WTgeneral}
\frac{\langle W[L]T_{11}( {y}) \rangle}{\langle W[L] \rangle}&=&-\frac{h_W}{r^3}\,,\\\nonumber
\frac{\langle W[L]T_{ \kappa \sigma}( {y}) \rangle}{\langle W[L] \rangle}&=&-\frac{h_W}{r^3}(3n_{\kappa}n_{\sigma}-2 \delta_{\kappa\sigma})\,,\\\nonumber
\frac{\langle W[L]T_{ 1\kappa}( {y}) \rangle}{\langle W[L] \rangle}&=&0\,,
\end{eqnarray}
where $\kappa, \sigma=2, 3$ and we have defined
\begin{equation}
r=\sqrt{y_2^2+y_3^2}\,, \qquad n_\kappa=\frac{y_\kappa}{r}\,.
\end{equation}
Dynamical information is contained in $h_W$, which is a non-trivial function of $N$ and $k$.

\section{Holographic computations}
\label{sec:CorrHolo}
In this section, we compute the Wilson line/loop OPE coefficients by holography. According to the AdS/CFT correspondence, ABJM theory with large $N$ and finite $k$ is dual to M-theory on $AdS_4\times S^7/{ \mathbb{Z}_k}$. Our calculation captures the leading-order result in the large $N$ limit while keeping $k$ fixed. The eleven-dimensional supergravity background consists of the metric
\begin{eqnarray}
ds^2=g_{MN}dX^MdX^N=ds^2_{AdS_4}+ 4 ds^2_{S^7/{ \mathbb{Z}_k}}\,,
\end{eqnarray}
and the four-form flux
\begin{equation}
F_4=3\Omega_{AdS_4}\,
\end{equation}
where we have set the radius of the $AdS_4$ to $1$, and $\Omega_{AdS_4}$ denotes the volume form of the unit $AdS_4$.
From the AdS/CFT dictionary, the eleven-dimensional Planck length is given by
\begin{equation}
l_p=\left(\frac{2}{\pi^2kN}\right)^{1/6}\,.\label{adscft}
\end{equation}

\paragraph{Euclidean $AdS_4$ metric} The metric on the unit (Euclidean) $AdS_4$ ($EAdS_4$) in Poincar\'e coordinates is
\begin{equation}
ds^2_{AdS_4}=\frac{1}{z^2}(dz^2+\delta_{ij}dx^idx^j)\,,
\end{equation}
where $i, j=1, 2, 3$.

\paragraph{Parametrization of $S^7/{ \mathbb{Z}_k}$} To describe the metric on $S^7/{ \mathbb{Z}_k}$, we parameterize the space using four complex coordinates $z^I, (I=1, \cdots, 4)$ subject to the constraint  $\sum_{I=1}^4|z^I|^2=1$. Explicitly,
\begin{eqnarray}
z^1&=&\cos\xi \cos \frac{\theta_1}{2} \exp\left[\mathrm{i} (\zeta+\frac{\psi+\varphi_1}{2})\right]\,,\\
z^2&=&\cos\xi \sin \frac{\theta_1}{2} \exp \left[\mathrm{i} (\zeta+\frac{\psi-\varphi_1}{2})\right]\,,\\
z^3&=&\sin\xi \cos \frac{\theta_2}{2} \exp \left[\mathrm{i} (\zeta+\frac{-\psi+\varphi_2}{2})\right]\,,\\
z^4&=&\sin\xi \sin \frac{\theta_2}{2} \exp \left[\mathrm{i} (\zeta+\frac{-\psi-\varphi_2}{2})\right]\,,
\end{eqnarray}
with the coordinate ranges:
\begin{align}
0\leq \xi \leq \frac{\pi}{2},\quad -\pi \leq \psi \leq \pi,\quad 0\leq \theta_i \leq \pi,\quad 0\leq \varphi_i \leq 2\pi,\quad (i=1, 2),
\end{align}
and the angular identification $\zeta \sim \zeta+\frac{2\pi}{k}$ due to the $ \mathbb{Z}_k$ quotient.
The metric on $S^7/\mathbb{Z}_k$, induced from the flat metric $ds_{ \mathbb{C}^4}=\sum_{i=1}^4 dz^Id\bar{z}_I$, takes the form
\begin{equation}
ds_{S^7/{ \mathbb{Z}_k}}=(d\zeta+A)^2+ds_{\mathbb {CP}^3}\,,
\end{equation}
where the one-form $A$ is given by
\begin{equation}
A=\frac{1}{2}\cos(2\xi) d\psi+\frac{1}{2}\cos^2\xi \cos\theta_1 d\varphi_1+\frac{1}{2}\sin^2\xi \cos\theta_2 d\varphi_2\,,
\end{equation}
and the Fubini-Study metric on $\mathbb{CP}^3$ is
\begin{eqnarray}
ds_{\mathbb{CP}^3}&=&d\xi^2+\frac{1}{4}\cos^2\xi (d\theta_1^2+\sin^2\theta_1 d\varphi_1^2)
+\frac{1}{4}\sin^2\xi (d\theta_2^2+\sin^2\theta_2 d\varphi_2^2)\nonumber\\
&+&\cos^2\xi \sin^2\xi (d\psi+\frac12 \cos\theta_1 d\varphi_1-\frac12 \cos\theta_2 d\varphi_2)\,.
\end{eqnarray}
\paragraph{Four-Form Flux} In Poincar\'e coordinates, the  four-form flux $F_4$ and its corresponding three-form potential $c_3$ are
\begin{align}
F_4=&\,\frac{3}{z^4}dz\wedge dx^1\wedge dx^2\wedge dx^3\,,\\\nonumber
    c_3=&\,-\frac{1}{z^3}dx^1\wedge dx^2\wedge dx^3\,.
\end{align}

\subsection{Holographic description of the operators }
We now employ the M-theory framework to compute the normalized correlation function of a Wilson loop operator $W[L]$ or $W[C]$ and a local operator. This framework is dual to ABJM theory in the large $N$ limit with $k$ fixed. Notice that this is different from the 't~Hooft limit  which is $N, k\to \infty$ with their ratio $N/k$, known as the 't~Hooft coupling $ \lambda$, fixed.

\paragraph{The Wilson line} In the holographic dual, the WL is described by a probe M2-brane embedded in the eleven-dimensional supergravity background $AdS_4\times S^7/{ \mathbb{Z}_k}$. The bosonic part of the M2-brane action consists of the Dirac-Born-Infeld (DBI) term and the Wess-Zumino (WZ) term~\footnote{Since we only use the bosonic part, it will be referred to as the membrane action or M2-brane action in the following.}:
\begin{equation}
S_{\text{M2}}=T_{\text{M2}}\left(\int d^3\sigma \sqrt{\mathrm{det}\tilde{g}}-P[c_3]\right)\,
\end{equation}
where $\tilde{g}_{ab}=\partial_a X^M \partial_b X^N g_{MN}$ is the induced metric on the M2-brane worldvolume and $P[c_3]$ is the pullback of $c_3$ to the membrane worldvolume. Here $T_{\text{M2}}$ is the tension of the M2-brane
\begin{equation}\label{tm2}
T_{\text{M2}}=\frac{1}{4\pi^2 l_p^3}\,.
\end{equation}
For the computation of WL one-point functions in this paper, the Wess-Zumino term does not contribute, as the M2-brane wraps two directions in $AdS_4$ and one direction in $S^7/{ \mathbb{Z}_k}$.\par

The holographic description of $W[L]$ (in the  fundamental representation of the supergroup $U(N|N)$) is in term of the following probe M2-brane in $AdS_4\times S^7/{\mathbb {Z}}_k$. Denoting the worldvolume coordinates of this M2-brane as $(\sigma^0, \sigma^1, \sigma^2)$, the embedding of this brane is
\begin{eqnarray}\label{embedding1} && z=\sigma^0\,, x_1=\sigma^1\,, x_2=x_3=0\, ,\nonumber\\
&& \zeta=\sigma^2+\zeta^0\,, \xi=\xi^0\,, \theta_1=\theta_1^0\,, \varphi_1=\varphi_1^0\,, \theta_2=\theta_2^0\,, \varphi_2=\varphi_2^0\,, \psi=\psi^0\,, \end{eqnarray}
where  $\sigma^0\ge 0, \sigma^1\in  (-\infty, \infty), \sigma^2\in [0, \frac{2\pi}{k}]$, and $\zeta^0, \xi^0, \theta_1^0, \varphi_1^0, \theta_2^0, \varphi_2^0, \psi^0$ are constants. The worldvolume of this M2-brane has the the topology $AdS_2\times S^1$ with $AdS_2$ in $AdS_4$ and $S^1$ along the $\zeta$-cycle. Notice that the above $AdS_2$ is in the Poincar\'e coordinates,
 and the $\zeta$-cycle is the M-theory circle when we consider reduction of  the M-theory on $AdS_4\times S^7/{ \mathbb{Z}}_k$ to the IIA superstring theory on $AdS_4\times  \mathbb{CP}^3$. The parameters $\alpha_I$  defining the Wilson line in the boundary theory are related to the brane embedding via~\cite{Lietti:2017gtc}
\begin{equation}\alpha_I=\bar{z}_I(\xi^0, \theta_1^0, \varphi_1^0, \theta_2^0, \varphi_2^0, \zeta^0)\,.\end{equation}  

\paragraph{The circular Wilson loops} The M2-brane dual to circular Wilson loops has worldvolume $EAdS_2\times S^1$ with  $EAdS_2$ now in the global coordinates.   Let us denote the worldvolume coordinates of the M2-brane by $(\sigma^0, \sigma^1, \sigma^2)$. The map $X^m(\sigma)$ from the membrane worldvolue to the $AdS_4$ part of the background geometry is specified by:
\begin{eqnarray}
\label{embedding2} 
&& z=R \sin \sigma^1\,, \quad x_1=R \cos \sigma^1\cos\sigma_0\,, \quad x_2=R\cos\sigma^1 \sin\sigma^0\,,\quad x_3=0\,,
\end{eqnarray}
The map $X^\alpha(\sigma)$ from the worldvolume to the $S^7/\mathbb{Z}_k$ part and the relation between the embedding and the parameters $\alpha_I$ is the same as the previous Wilson line case.
where the coordinate ranges are
\begin{align}
0\leq \sigma^0\leq 2\pi,\quad 0\leq \sigma^1\leq \frac{\pi}{2},\quad  0\leq \sigma^2 \leq \frac{2\pi}{k}\,,
\end{align}

\paragraph{The local operators} In the holographic framework, the local operators we consider correspond to perturbations of the background metric and flux fields.\par

According to the $AdS_4/CFT_3$ duality, the boundary stress tensor $T_{\mu\nu}$ is dual to metric fluctuations in the $AdS_4$ subspace of the full $AdS_4\times S^7/ \mathbb{Z}_k$ background. The perturbed metric takes the form
\begin{equation}\label{eq:varg}
G_{mn}=g_{mn}+\delta g_{mn}\,,
\end{equation}
where $g_{mn}$ is the unperturbed $AdS_4$ metric. The explicit expression for $\delta g_{mn}$ will be provided in the next subsection when we perform concrete calculations.\par

The chiral primary operators $\mathcal{O}^A$ are dual to fluctuations of both the metric and the three-form potential in the eleven-dimensional background:
\begin{eqnarray}
    G_{MN}&=&g_{MN}+\delta g_{MN}^A \mathbf{Y}^A\,,\label{eq:gmn}\\
    C_{MNP}&=&c_{MNP}+\delta c_{MNP}^A \mathbf{Y}^A\,,\label{eq:cmnp}
\end{eqnarray}
where $\mathbf{Y}^A=(C^A)_{I_1\cdots I_L}^{J_1\cdots J_L} \alpha_{J_1}\cdots \alpha_{J_L} \bar{\alpha}^{I_1} \cdots \bar{\alpha}^{I_L}$ are spherical harmonics of ${S}^7/\mathbb {Z}_k$ with radius $1$.  The explicit form of the fluctuations will be given when we carry out explicit computations in the following sections.

\subsection{Holographic calculations of $\langle W[L] T_{\mu\nu}\rangle/\langle  W[L]\rangle$}
We compute the variation of the M2-brane action, $\delta S_{\text{M2}}$, induced by fluctuations of the background metric in Eq.~\eqref{eq:varg}. From this, we derive the correlation function between the Wilson loop $W[L]$ and the stress tensor $T_{\mu\nu}(y)$ as~\footnote{The factor $2$ in the Eq.~\eqref{correlator} was explained in~\cite{Georgiou:2025mgg}.}

\begin{eqnarray}
\label{correlator}
\frac{\langle W[L]T_{\mu\nu}( {y}) \rangle}{\langle W[L] \rangle}&=&-2\frac{\delta S_{\text{DBI}}}{\delta \hat{h}^{\mu\nu}( {y})}\,.
\end{eqnarray}

Here a key comment is in order. As stressed in~\cite{Bajnok:2014sza, Yang:2021kot}, in holographic computations of three-point functions involving classical string/brane solutions, we should average over classical configurations of the strings/branes. This is equivalent to averaging over the orbit generated by the broken global symmetries acting on the strings/branes. However, in the case at hand and that considered below, the membranes (and the dual Wilson loops) are coherent states instead of eigenstates with respect to the global charges broken by the membrane solutions, in the sense of~\cite{Bajnok:2014sza}. So no orbit average is needed here.

We now go back to the concrete calculations.
The explicit form of the metric fluctuation $\delta g_{mn}$ in \eqref{eq:varg}  is given by~\cite{Liu:1998bu}~\footnote{The indices of $J_{mn}$ and  $\mathcal{P}_{\mu\nu,\, \rho\sigma}$ are always contracted with the flat metric. }
\begin{eqnarray}\label{fluctuations}
\delta g_{mn}(z, x)
&=&\kappa_3z\int d^3y \frac{1}{f^3} J_{ m\mu}(X-Y)\, J_{ n\nu} (X-Y)\, \mathcal{P}^{\mu\nu, \,\rho\sigma}\hat{h}_{\rho\sigma}( {y})\,,
\end{eqnarray}
where $X=(z, x)$ denotes a bulk point in $AdS_4$ and $Y=(0, y)$ is a point at the  boundary. $\hat{h}_{\rho\sigma}( {y})$ is the fluctuation of the metric on the boundary of $AdS_4$ which couples to the stress tensor $T^{\rho\sigma}(y)$. The functions $f, J_{mn}, \mathcal{P}_{\mu\nu, \rho\kappa}$ are defined as~\footnote{The definition of $X_m$ is $X_m:=\delta_{mn}X^n$.}
\begin{eqnarray}
f&:=&|X-Y|^2:=z^2+( {x}- {y})^2\,,\\
J_{mn}(X)&:=&\delta_{mn} -2\frac{X_m X_n}{|X|^2}\,,\\
\mathcal{P}^{\mu\nu, \,\rho\sigma}&:=&\frac{1}{2}(\delta^{\mu\rho}\delta^{\nu\sigma}+\delta^{\mu\sigma}\delta^{\nu\rho})-\frac{1}{3}\delta^{\mu\nu}\delta^{\rho\sigma}\,,
\end{eqnarray}
The constant $\kappa_d$ for general $d$ is given by
\begin{equation}
\kappa_d=\frac{d+1}{d-1}\frac{\Gamma[d]}{\pi^{d/2}\Gamma(d/2)}\,,
\end{equation}
yielding $\kappa_3=8/{\pi^2}$ (corresponding to $d=3$) for the case at hand. The induced metric components are
\begin{eqnarray}
\tilde{g}_{00}=\frac{1}{({\sigma^0})^2}\,,\qquad 
\tilde{g}_{11}=\frac{1}{({\sigma^0})^2}\,,\qquad
\tilde{g}_{22}=4\,.
\end{eqnarray}
Expanding Eq.~\eqref{fluctuations} explicitly, we obtain
\begin{eqnarray}
\delta g_{zz}&=&\frac{8}{\pi^2}\int d^3y \frac{z}{f^3} \left[\delta _{\mu z} -\frac{2z(x-y)_\mu}{f}\right] \left[\delta _{\nu z} -\frac{2z(x-y)_\nu}{f}\right]\nonumber\\
&&\qquad\times\left(\frac{1}{2}\delta^{\mu\rho}\delta^{\nu\sigma}+\frac{1}{2}\delta^{\mu\sigma}\delta^{\nu\rho}-\frac{1}{3}\delta^{\mu\nu}\delta^{\rho\sigma}\right)\hat{h}_{\rho\sigma}( {y})\nonumber\\
&=&\frac{32}{\pi ^2}\int d^3y \frac{z^3(x-y)_\mu(x-y)_\nu}{f^5}\left(\frac{1}{2}\delta^{\mu\rho}\delta^{\nu\sigma}+\frac12\delta^{\mu\sigma}\delta^{\nu\rho}-\frac{1}{3}\delta^{\mu\nu}\delta^{\rho\sigma}\right)\hat{h}_{\rho\sigma}( {y})\,,
\end{eqnarray}
and
\begin{eqnarray}
\delta g_{x^1x^1}&=&\frac{8}{\pi^2}\int d^3y \frac{z}{f^3} \left[\delta_{\mu x^1} -\frac{2(x-y)_1(x-y)_\mu}{f}\right] \nonumber\\
&\times&\left[\delta_{\nu x^1} -\frac{2(x-y)_1(x-y)_\nu}{f}\right]\left(\frac{1}{2}\delta^{\mu\rho}\delta^{\nu\sigma}+\frac12\delta^{\mu\sigma}\delta^{\nu\rho}-\frac{1}{3}\delta^{\mu\nu}\delta^{\rho\kappa}\right) \hat{h}_{\rho\sigma}( {y})\,.
\end{eqnarray}
Now we proceed to compute different components of the correlation function \eqref{correlator}.
\begin{itemize}
\item For $\mu=\nu=1$, the relevant metric variations are
\begin{eqnarray*}
\int d^3\sigma\frac{ \delta g_{zz}}{\delta \hat{h}_{11}(y)}
=\frac{32}{\pi^2} \int d^3\sigma \frac{z^3}{f^5}\left(\frac 23 x_1^2 -\frac{1}{3}r^2\right)
\end{eqnarray*}
and
\begin{eqnarray*}
\int d^3\sigma\frac{\delta g_{x_1x_1}}{\delta \hat{h}_{11}(y)} =\frac{8}{\pi^2} \int d^3\sigma \frac{z}{f^3}\left[\left(1-\frac{2x_1^2}{f}\right)^2\right.-\frac{1}{3}\left(1-\frac{4x_1^2}{f} \right.
+\left.\left.\frac{4x_1^2 (x_1^2 +r^2)}{f^2}\right)\right]\,.
\end{eqnarray*}
where $ {y}= (0 , y_2 , y_3)$, and we have defined $r:=\sqrt{y_2^2+y_3^2} , n_\kappa:={y_\kappa}/{r}$, for $\kappa=2, 3$. The correlation function evaluates to
\begin{eqnarray}
\frac{\langle W[L]T_{11}( {y}) \rangle}{\langle W[L] \rangle}
&=& -\frac{128}{\pi^2}T_{\text{M2}}\int_{0}^{\frac{2\pi}{k}} d\zeta \int_{0}^{\infty} dz \int_{-\infty}^{\infty} dx_1 \frac{z[r^4-4r^2 x_1^2 +(x_1^2+z^2)^2]}{12(r^2+x_1^2+z^2)^5}\nonumber\\
&=& -\frac{T_{\text{M2}}}{kr^3} =-\frac{1}{4\sqrt{2}\pi} \sqrt{\frac{k}{N}} \frac{1}{r^3}\,,
\end{eqnarray}
where we have used \eqref{adscft} and \eqref{tm2}. 
\item For $\mu=1$, $\nu=2,3$, the metric variations vanish:
\begin{equation}
  \int d^3\sigma\frac{\delta g_{zz}}{\delta \hat{h}^{1\nu}(y)} = \int d^3 \sigma\int d^3 y \frac{-4z^3 x_1 y^\nu}{f^5}=0\,,
\end{equation}
and
\begin{equation}
\int d^3\sigma\frac{\delta g_{x_1x_1}}{\delta \hat{h}^{1\nu}(y)} =\int d^3\sigma \int d^3 y\frac{z}{f^3}\left(1-\frac{2x_1^2}f\right)\frac{2x_1y^\nu}{f}=0\,.
\end{equation}
Thus, the corresponding components of Eq.~\eqref{correlator} are zero.

\item For $\mu,\nu=2,3$, the metric fluctuations take the form
\begin{align}
\delta g_{zz} &= \int d^3 y \frac{32 z^3}{\pi^2 f^5} \left[ r^2 \left( n^\mu n^\nu - \frac{1}{3} \delta^{\mu\nu} \right) - \frac{1}{3} x_1^2 \delta^{\mu\nu} \right] \hat{h}_{\mu\nu}(y)\,,\\\nonumber
\delta g_{x^1 x^1} &= -\int d^3 y \frac{8z}{3\pi^2 f^3} \delta^{\mu\nu} \left[ 1 - \frac{4x_1^2}{f} + \frac{4x_1^2 (x_1^2 + r^2)}{f^2} \right] \hat{h}_{\mu\nu}(y)\,,
\end{align}
leading to the correlation function
\begin{equation}
\frac{\langle W[L] T_{\mu\nu}(y) \rangle}{\langle W[L] \rangle} = -2\frac{\delta S_{\text{M2}}}{\delta \hat{h}^{\mu\nu}(y)} =- \frac{1}{4\sqrt{2}\pi} \sqrt{\frac{N}{k}} \frac{1}{r^3} \left( 3n^\mu n^\nu - 2 \delta^{\mu\nu} \right)\,.
\end{equation}
\end{itemize}
The results for the correlation function~(\ref{correlator}) can be summarized  as
\begin{eqnarray}
\frac{\langle W[L]T_{11}( {y}) \rangle}{\langle W[L] \rangle}&=&-\frac{h_W}{r^3}\,,\\
\frac{\langle W[L]T_{ \kappa \sigma}( {y}) \rangle}{\langle W[L] \rangle}&=&-\frac{h_W}{r^3}(3n_{\kappa}n_{\sigma}-2 \delta_{\kappa\sigma})\,,\\
\frac{\langle W[L]T_{ 1\kappa}( {y}) \rangle}{\langle W[L] \rangle}&=&0\,,\end{eqnarray}
with 
\begin{equation}
\label{eq:scalinghW}
h_W=\frac{1}{4\sqrt{2}\pi}\sqrt{\frac{N}{k}}=\frac{\sqrt{\lambda}}{4\sqrt{2}\pi}\,,
\end{equation}
and $\kappa, \sigma=2, 3$. 
This structure agrees with symmetry considerations \eqref{eq:WTgeneral}, with the dynamical information encoded in the scaling function $h_W$.

\subsection{Holographic calculations of $\langle \mathcal{O}\rangle_{W[C]}$}
In this subsection, we compute the normalized correlation function of the half-BPS circular Wilson loop $W[C]$ and the $1/3$-BPS chiral primary operator $\mathcal{O}^A$

\begin{equation}
\langle \mathcal{O}^A (y) \rangle_{W[C] }:=\frac{\langle \mathcal{O}^A(y) W[C] \rangle }{\langle W[C] \rangle \sqrt{\mathcal{N}_{\mathcal{O}^A}}}\,,
\end{equation}
where the normalization factor $\mathcal{N}_{\mathcal{O}^A}$ is defined via the two-point function of ${\mathcal{O}^A}$

\begin{equation}
\langle \mathcal{O}^A( {x}) \mathcal{O}^{B\dagger}( {y}) \rangle=\frac{\delta^{AB}\mathcal{N}_{\mathcal{O}^A}}{|x-y|^{2\Delta_A}}\,,
\end{equation}
with $\Delta_A$ being the conformal dimension of $\mathcal{O}^A$. We will focus on the OPE limit ${L}\gg R$ where $L$ is the distance between the local operator and the center of the Wilson loop.\par

The CPO corresponds to fluctuations of the background fields in Eqs.~\eqref{eq:gmn} and \eqref{eq:cmnp}, with explicit forms given by~\cite{Biran:1983iy, Castellani:1984vv, Bastianelli:1999bm, Drukker:2008jm}:
\begin{eqnarray}
\label{fl}
   \delta g_{mn}^A &=& \frac{4}{J+2}\left[\nabla_m\nabla_n +\frac{J(J+6)}{8}g_{mn}\right] s^A-\frac{7J}{6}g_{mn} s^A\,,\\
        \delta g_{\alpha\beta}^A&=& \frac{1}{3}Jg_{\alpha\beta} s^A\,,\\
       \delta c^A_{mnp} &=& 2\epsilon_{mnpq } \nabla^{q} s^A\,,
\end{eqnarray}
 where $J=2 \Delta=2L$. 
The scalar field $s^A(x, z)$ depends only on the coordinates $(x, z)$ of $AdS_4$ and  is determined by the boundary source $s_0^A(y)$
\begin{equation}
    s^A( {x},z) = \int d^3{y}G_{\Delta}( {y}; {x},z)s_0^A( {y})\,,
\end{equation}
where $G_\Delta( {y};  {x},z)$ is the boundary-to-bulk propagator
\begin{equation}
    G_\Delta( {y};  {x},z)=c\left(\frac{z}{z^2+| {x-y}|^2}\right)^\Delta\,,
\end{equation}
with the constant $c$ being~\cite{Drukker:2008jm}
\begin{equation}\label{constant}
c=2^{\Delta-1}\pi l_p^4 \sqrt{k} \frac{\Delta+1}{\Delta}\sqrt{2\Delta+1}\,.
\end{equation}
In the OPE limit ${L}\gg R$, we can approximate
\begin{eqnarray}
G_{\Delta}( y; {x},z)&\approx& c\frac{z^{\Delta}}{\tilde{L}^{2\Delta}}\,,\\
\partial_z s^A&\approx&\frac{\Delta}{z}s^A\,,\\
\partial_{\mu} \partial_{\nu }s^A&\approx& \delta_{\mu}^z \delta_{\nu}^z \frac{\Delta(\Delta -1)}{z^2}s^A ,
\end{eqnarray}
 For Euclidean $AdS_4$ in the Poincar\'e coordinates,  the Christoffel symbols are
\begin{eqnarray}
    \Gamma^z_{mn}=\frac{1}{2}g^{zz}(\partial_{m} g_{z n } +\partial_{n} g_{z m} -\partial_z g_{m n})
    = z g_{mn}-\frac2{z}\delta^z_m \delta^z_n\,.
\end{eqnarray}
Substituting these into \eqref{fl}, the metric fluctuation simplifies to:
\begin{eqnarray}
    \delta g_{mn}^A
    =-\frac{2}{3}J g_{mn}s^A + \frac{J}{z^2}\delta_{m}^z \delta_{n}^z s^A\,.
\end{eqnarray}
The correlation function is computed from the variation of the M2-brane action due to background field fluctuations. As mentioned above, the Wess-Zumino term does not contribute, so the action variation is 
\begin{equation}
    \delta S_{\text{M2}} =T_{\text{M2}}\int d^3\sigma \sqrt{\mathrm{det}\tilde{g}_{ab}}\frac{1}{2}\tilde{g}^{ab}\partial_a X^M \partial_b X^N \delta g_{MN}^A\mathbf{Y}^A\,,
\end{equation}
where the induced metric $\tilde{g}_{ab}=\partial_a X^M \partial_b X^N g_{MN}$ has components: 
\begin{eqnarray}
    \tilde{g}_{00}=\cot^2\sigma^1\,,\qquad
    \tilde{g}_{11}=\sec^2\sigma^1\,,\qquad
    \tilde{g}_{22}=4\,.
\end{eqnarray}
The contraction yields,
\begin{eqnarray}
    \tilde{g}^{ab}\partial_a X^M \partial_b X^N \delta g_{MN}^A&=& -\frac{2}{3}Js^A -\frac{2}{3}Js^A +J \tilde{g}^{11}\frac{(\partial_1 z)^2}{z^2}s^A +\frac{1}{3}Js^A\nonumber\\
  &=& -J\sin^2 \sigma_1 s^A\,,
\end{eqnarray}
leading to the action variation
\begin{eqnarray}
    \delta S_{\text{M2}}&=&T_{\text{M2}}\int d^3 \sigma \left(2\frac{\cos \sigma_1}{\sin^2 \sigma_1}\right)\frac{1}{2}(-J\sin^2 \sigma) s^A\mathbf{Y}^A\nonumber\\
    &=&-JT_{\text{M2}}\int d^3\sigma \cos \sigma_1 s^A\mathbf{Y}^A\,.
\end{eqnarray}
Using the AdS/CFT dictionary, the correlation function reads
\begin{eqnarray}
    \frac{\langle W[C] \mathcal{O}^A( {y}) \rangle }{\langle W[C] \rangle \sqrt{\mathcal{N}_\mathcal{O}}}&=&-\frac{\delta S_{\text{DBI}}}{\delta s_0^A( {y})}
    =JT_{\text{M2}}\int d^3\sigma \cos \sigma_1 G_\Delta( {y}; {x},z)\mathbf{Y}^A\nonumber\\
&=&\frac{2^{\Delta +\frac{1}{4}}\sqrt{\pi(2\Delta +1)}}{k\lambda^{\frac{1}{4}}}\frac{R^{\Delta}}{{L}^{2\Delta}}\mathbf{Y}^A\,,
\end{eqnarray}
where we have used \eqref{adscft}, \eqref{tm2} and \eqref{constant}. We define the OPE coefficients $C_{\mathcal{O}^A}$ as 
\begin{equation}
    \frac{\langle W[C] \mathcal{O}^A( {y}) \rangle }{\langle W[C] \rangle\sqrt{\mathcal{N}_\mathcal{O}}}
=C_{\mathcal{O}^A}\frac{R^\Delta}{L^{2\Delta}}\mathbf{Y}^A,
\end{equation}
and obtain
\begin{equation}
C_{\mathcal{O}^A}=\frac{2^{\Delta +\frac{1}{4}}\sqrt{\pi(2\Delta +1)}}{k\lambda^{\frac{1}{4}}}\,.
\end{equation}

\section{Field theory results from localization}
\label{sec:localization}
\subsection{The case of $\langle W[L] T_{\mu\nu}\rangle/\langle W[L] \rangle $}
In this subsection, we compare our holographic computation of $\langle W[L] T_{\mu\nu}\rangle/\langle W[L]\rangle$ with field theory result. Symmetry fixes this correlation function to the form \eqref{eq:WTgeneral} up to a scaling function $h_W(k, N)$, which is known to be related to the Bremsstrahlung function $\mathcal{B}(k, N)$ by~\footnote{This is first conjectured in \cite{Lewkowycz:2013laa}  for general spacetime dimensions and proved for four-dimensional ${\mathcal{N}}=2$  theories~\cite{Bianchi:2018zpb}.}
\begin{equation}
h_W(k, N) = \frac{\mathcal{B}(k, N)}{2}.
\end{equation}
The Bremsstrahlung function for ABJM theory has been determined non-perturbatively in M-theory limit~\cite{Beccaria:2025vdj} based on~\cite{Klemm:2012ii},  and its large $N$ expansion (with finite $k$) takes the form
\begin{equation}
\label{eq:expandB}
\mathcal{B}(k, N) \big|_{N \gg 1} = \frac{1}{2\pi} \sqrt{\frac{N}{2k}} - \frac{1}{2\pi k}\cot\frac{2\pi}{k} + \mathcal{O}(N^{-1/2}).
\end{equation}
Consequently, the scaling function $h_W(N, k)$ in the M-theory limit is given by
\begin{equation}
h_W(k, N) \big|_{N \gg 1} = \frac{\sqrt{N}}{4\pi \sqrt{2k}} + \mathcal{O}(N^0).
\end{equation}
This matches precisely with the holographic result~\eqref{eq:scalinghW}.

\subsection{The case of  $\langle \mathcal{O}\rangle_{W[C]}$}\label{subsection:fieldWO}
In this subsection, we review the localization computation of the correlation function of a Wilson line and a CPO with conformal dimension $\Delta=1$ \cite{Guerrini:2023rdw} and compare it with the holographic result~\footnote{As we will show below, there is  simple relation between the OPE coefficients of  the Wilson line and the Wilson loop. }. Following \cite{Guerrini:2023rdw}, we consider the ABJM theory on $\mathbb{R}^3$ with Euclidean signature. The half-BPS Wilson line is placed along the $x^3$-axis at $(x^1, x^2, x^3) = (0, 0, s)$ and the polarization vector $\alpha_I$ in the definition of the Wilson loop is chosen to be $\alpha_I=(1, 0, 0, 0)$~\footnote{The WL is put along $x^3$ direction, instead of the $x^1$ direction as in section~\ref{sec:corr1}. As a result, we should change the spinors~\eqref{eq:spinors} in the definition of the Wilson line accordingly for the computation.  However, since we are only concerned about OPE coefficients, which is spacetime independent, the changes of the conventions  do not affect the result.}. Local operators are inserted at $(r \cos \tau, r \sin \tau, 0)$, and their polarization vectors are given by~\footnote{We have adapted to our notation.}
\begin{eqnarray}
n_I=\frac{1}{\sqrt{2}}\left( e^{-\frac{\ri}{2}\tau}, 0, e^{\frac{\ri}{2}\tau}, 0\right)\,,\qquad
\bar{n}^I=\frac{1}{\sqrt{2}}\left(e^{\frac{\ri}{2}\tau}, 0, -e^{-\frac{\ri}{2}\tau}, 0 \right)\,.
\end{eqnarray}
The $1/3$-BPS operator with $\Delta=1$ in the translation-twisted frame along the circle is
\begin{equation}
\label{eq:CPOdelta1}
\mathcal{O}(\tau)=\mathrm{Tr}(n_I(\tau) Y^I(\tau) \bar{n}^J(\tau) \bar{Y}_J(\tau)) \,.
\end{equation}
In this frame, the propagator $d_{ij}$ takes the form
\begin{equation}
d_{12}=\frac{n(\tau_1)\cdot \bar{n}(\tau_2)}{|x(\tau_1)-x(\tau_2)|}=\frac{\ri \sin \frac{\tau_1-\tau_2}{2}}{2 r |\sin \frac{\tau_1-\tau_2}{2}|}\,,
\end{equation}
leading to
\begin{equation}
d_{12}d_{21}=\frac1{4r^2}\,,
\end{equation}
which is independent of $\tau_i$. The two-point function of $\mathcal{O}$ is normalized as
\begin{equation}
\label{two-point}
\langle  \mathcal{O}(\tau_1)\mathcal{O}(\tau_2)\rangle=\mathcal{N}_\mathcal{O} d_{12}d_{21}\,,
\end{equation}
where $\mathcal{N}_\mathcal{O}$ is the normalization constant.\par

\paragraph{Localization calculation} For arbitrary finite $N$ and $k$, the integrated two-point function is given by~\cite{Agmon:2017xes, Binder:2019mpb}
\begin{eqnarray}
\left\langle \int d\tau_1 \mathcal{O}(\tau_1) \int d\tau_2 \mathcal{O}(\tau_2)\right\rangle=-\frac{1}{16\pi^2 r^4} \frac1Z \frac{\partial^2 Z[m]}{\partial m^2}\Bigg|_{m=0}\,,
\end{eqnarray}
where $Z[m]$ is the partition function of the mass-deformed ABJM theory where $m$ is the mass-deformation parameter. Using localization it can be expressed as~\cite{Kapustin:2009kz,Kapustin:2010xq, Jafferis:2010un, Hama:2010av}
\begin{eqnarray}
Z[m]&=&\frac1{(N!)^2(2\pi)^{2N}}\int d\lambda_i d\mu_i \exp\left[\frac{\ri k}{4\pi}(\sum \lambda_i^2-\sum \mu_i^2)\right]\\
&&\times \frac{\prod_{i\neq j}^{N} 2 \sinh \frac{\lambda_i-\lambda_j}2 \prod_{i\neq j}^N 2 \sinh \frac{ \mu_i-\mu_j}2}{\prod_{i, j}2 \cosh\frac{\lambda_i-\mu_j}2 2\cosh (\frac{\mu_j-\lambda_i }2-\pi m r)}\,.
\end{eqnarray}
Using Fermi gas approach, all order perturbatrive $1/N$ corrections to  $Z[m]$ can be resumed~\footnote{We omit all instanton corrections. }. The result is given in terms of the  Airy function~\cite{Marino:2011eh, Guerrini:2023rdw}
\begin{equation}
Z^{\mathrm{pert.}}[m]=e^A C^{-1/3} \mathrm{Ai}[C^{-1/3}(N-B)]\,,
\end{equation}
where $A, B, C$ are functions of $k, r, m$. Here we do not need the explicit expressions for $A$ and $B$. As for $C$, we have 
\begin{equation}
    C=\frac{2}{\pi^2 k (1+4 m^2 r^2)}\,.
\end{equation}

 Using the asymptotic expansion of the Airy function
\begin{equation}
\mathrm{Ai}(x)\sim\exp\left(-\frac{2}{3}x^{3/2}\right)\,,
\end{equation}
where an irrelevant factor $(1/x)^{1/4}/(2\sqrt{\pi})$ has been omitted, the large $N$ limit of the integrated two-point function reads
\begin{eqnarray}
\label{eq:OO1}
\left\langle \int d\tau_1 \mathcal{O}(\tau_1) \int d\tau_2 \mathcal{O}(\tau_2)\right\rangle&=&-\frac{1}{16\pi^2 r^4} \frac{1}{Z[m] }\frac{\partial^2 Z[m]}{\partial m^2}\Bigg|_{m=0}=\frac{\sqrt{2k}N^{3/2}}{12\pi r^2}\,.
\end{eqnarray}
Performing the $\tau$-integrals on both sides of \eqref{two-point} yields
\begin{eqnarray}
\label{eq:OO2}
\left\langle \int d\tau_1 \mathcal{O}(\tau_1) \int d\tau_2 \mathcal{O}(\tau_2)\right\rangle=\frac{\pi^2 \mathcal{N}_\mathcal{O}}{r^2}.
\end{eqnarray}
Comparing \eqref{eq:OO1} and \eqref{eq:OO2} fixes the normalization
\begin{equation}
\mathcal{N}_{\mathcal{O}}=\frac{\sqrt{2k}N^{3/2}}{12\pi^3}\,.
\end{equation}
The correlation function of the Wilson line $W[L]$ and $\mathcal{O}$ is given by
\begin{equation}
\frac{\langle W[L] \mathcal{O}(0) \rangle}{\langle W[L]  \rangle}=\frac{\mathcal{B}}{r},
\end{equation}
where $\mathcal{B}$ is the Bremsstranhlung function \eqref{eq:expandB}. Including the normalization factor, we have
\begin{equation}
\frac{\langle W[L] \mathcal{O}(0) \rangle}{\sqrt{\mathcal{N}_\mathcal{O}}\langle W[L]  \rangle}=\frac{\mathcal{B}}{\sqrt{\mathcal{N}_O}r}\,,
\end{equation}
The coefficient of $1/r$ is
\begin{equation}
\label{eq:resCPO}
\frac{\mathcal{B}}{\sqrt{\mathcal{N}_{\mathcal{O}}}}=\frac{\sqrt{3\pi }}{2^{3/4}k\lambda^{1/4}}\,,
\end{equation}
which is the prediction from localization.

\paragraph{Comparison with holography} The holographic computation of the Wilson loop OPE coefficient for the CPO \eqref{eq:CPOdelta1} gives
\begin{equation}
\mathcal{C}_{\mathcal{O}}=\frac{2^{5/4}\sqrt{3\pi}}{k\lambda^{1/4}},
\end{equation}
At first glance, this does not match \eqref{eq:resCPO}. However, note that the holographic correlation function for a normalized CPO of dimension $\Delta$ with the Wilson line takes the form
\begin{equation}
\label{eq:normalizedW}
\frac{\langle W[L]\mathcal{O}\rangle}{\sqrt{\mathcal{N}_\mathcal{O}}\langle W[L] \rangle}=\mathcal{C}_\mathcal{O} \mathbf{Y}^A \frac{1}{(2r)^\Delta},
\end{equation}
where $r$ is the distance from the CPO to the line, and $\mathbf{Y}^A$ is the spherical harmonic factor.

For the Wilson line considered here, the polarization vector is $\alpha_I = (1, 0, 0, 0)$, and the spherical harmonic evaluates to~\footnote{Recall that here the spherical harmonics is for $S^7/\mathbb{Z}_k$ with radius $1$.}
\begin{equation}
\mathbf{Y}^A=\bar{n}^I n_J \alpha_I \bar{\alpha}^J=1/2\,.
\end{equation}
Thus, the holographic prediction for $\Delta = 1$ is
\begin{equation}
\frac{\langle W[L]\mathcal{O}\rangle}{\sqrt{\mathcal{N}_\mathcal{O}}\langle W[L] \rangle}=\frac{\mathcal{C}_{\mathcal{O}}\mathbf{Y}^A}{2r}=\frac{\sqrt{3\pi}}{2^{3/4}k\lambda^{1/4} r}\,,
\end{equation}
which exactly matches the localization result \eqref{eq:resCPO}.\par

The factor of $2^\Delta$ in the denominator of \eqref{eq:normalizedW} arises from the near-line limit of the correlation function. For a circular Wilson loop of radius $R$ and a CPO at a generic point $(y_1, y_2, y_3)$, the correlation function takes the form in \eqref{eq:setupWO} \cite{Berenstein:1998ij, Alday:2011pf}
\begin{equation}
\frac{\langle W[C]\mathcal{O} \rangle} {\langle W[C] \rangle\sqrt{\mathcal{N}_\mathcal{O}} }=\tilde{C} \frac{R^\Delta}{((h^2+\rho^2-R^2)^2+4 R^2 h^2)^{\Delta/2}}\,,
\end{equation}
In the OPE limit $L = \sqrt{\rho^2 + h^2} \gg R$, this reduces to
\begin{equation}
\frac{\langle W[C]\mathcal{O} \rangle} {\langle W[C] \rangle\sqrt{\mathcal{N}_\mathcal{O}} }=\tilde{C} \frac{R^\Delta}{L^{2\Delta}}\,,
\end{equation}
Hence $\tilde{C}$ is identified with the OPE coefficient $\mathcal{C}_\mathcal{O}$.

For the Wilson line ($R \to \infty$), the correct limit is not obtained by naively sending $R\to\infty$, which is vanishing. Instead, we need to carefully send $r = \sqrt{(\rho - R)^2 + h^2} \to 0$ with $(\rho-R)/h$ finite. The geometric meaning of this limit is that when the operator is very near to the circle, it cannot distinguish between the circle and the straight line. This limit yields
\begin{equation}
\frac{\langle W[C] \mathcal{O} \rangle}{\langle W[C] \rangle \sqrt{\mathcal{N}_{\mathcal{O}}}} \approx \frac{{\mathcal{C}}_{\mathcal{O}}}{(2d)^\Delta}.
\end{equation}
This explains the factor of $2^\Delta$ in \eqref{eq:normalizedW}.

\section{Conclusion}
\label{sec:concl}
In this work, we have computed the correlation functions between a straight Wilson line (or a circular Wilson loop) and local BPS operators. Specifically, the local operator can be a stress energy tensor or a 1/3-BPS CPO in ABJM theory. By employing holographic methods, we derived explicit expressions for these one-point functions in the M-theory limit and compared them with predictions from supersymmetric localization. Our findings provide valuable strong coupling data for defect ABJM theory and establish a concrete basis for computing more general WL correlators holographically.\par

The matching between holographic and localization results underscores the power of combining complementary non-perturbative techniques in the study of defect CFTs. On the one hand, holography offers a geometric perspective, where WL correlators are mapped to the response of of string or brane configurations to certain fluctuations of the bulk background fields. On the other hand, localization reduces the problem to finite dimensional matrix integrals, enabling exact computations even at finite $N$ and $k$. There are many interesting directions one can pursue in the future.\par

Our work focused on half-BPS WLs in the fundament representation of the supergroup $U(N|N)$. A natural extension is to study WLs with less supersymmetries and/or  in higher-rank representations, such as symmetric or antisymmetric representations. For example, one can consider the $1/6$-BPS fermionic latitude WLs~\cite{Cardinali:2012ru} with a string dual given in~\cite{Aguilera-Damia:2014qgy}. However, in many cases the string description can be quite complex, or probe string/brane solutions are still lacking~\cite{Drukker:2019bev}. For bosonic 1/6-BPS WLs in the fundamental representation, the holographic description involves smearing the string worldsheet over $\mathbb{CP}^1$ inside $\mathbb{CP}^3$~\cite{Drukker:2008zx, Rey:1998ik}. For WLs in the antisymmetric representation, only the $1/6$-BPS D$6$-brane solution is known, with half-BPS ones remaining elusive. For instance, the string dual of fermionic $1/6$-BPS WLs~\cite{Ouyang:2015iza, Ouyang:2015bmy} in the fundamental representation of $U(N|N)$ involves complicated mixed boundary conditions on the worldsheet~\cite{Correa:2019rdk}. Extending our methods to these cases would yield new insights into the dependence of the defect correlator and its bulk duals on representation and supersymmetry~\footnote{In some cases, we would need to find the dual string/brane solutions first.}.

The computation of WL correlators lies at the intersection of multiple non-perturbative approaches, and further progress could be made by leveraging integrability and localization more broadly. While our work focused on specific BPS operators, a localization formula for generic CPO correlators would be highly desirable. Such a result could enable systematic studies of WL one-point functions beyond the stress energy tensor and 1/3-BPS CPOs with $\Delta=1$. In the planar limit, integrability techniques could be applied to compute WL one-point functions for non-BPS operators. For WLs that correspond to integrable boundary states, the OPE coefficients might be computable using the worldsheet $g$-function approach~\cite{Jiang:2019xdz, Jiang:2019zig}. Preliminary steps in this direction have been taken at weak coupling~\cite{Jiang:2023cdm}, but a strong-coupling analysis remains an open challenge.\par

While we focused on one-point functions, defect CFTs admit a richer class of observables, such as two-point functions of local operators in the presence of a WL, which encode defect operator spectra~\cite{Billo:2016cpy}. The computation of defect two-point functions are much more challenging as it depends on the conformal cross ratios in a non-trivial way. Recently, progress have been made in computing defect two-point functions in $\mathcal{N}=4$ SYM theory with defects of various co-dimensions by superconformal bootstrap~\cite{Gimenez-Grau:2023fcy, Chen:2025yxg}. It is very interesting to see whether such techniques can be generalized to the ABJM theory.

\acknowledgments

It is our pleasure to thank Bin~Chen, Yuezhou~Li, Hong~L\"u and Tadashi~Okazaki for very helpful discussions. The work of Y.J. is partly supported by Startup Funding no. 4007022326 of Southeast University. The work of X.-Y.~Z. and J.-B.~W. is supported  by the National Natural Science Foundation of China (NSFC) Grants No.~12375006, 11975164,  12247103, 11935009, and 
Tianjin University Self-Innovation Fund Extreme Basic Research Project Grant No.~2025XJ21-0007. 
This work is also supported by the NSFC Grant No.~12247103.

\bibliographystyle{JHEP}
\bibliography{WL}

\end{document}